\documentclass[final,3p,times,twocolumn]{elsartm}
\biboptions{comma,sort&compress}
\usepackage{graphicx}

\newcommand{\DP}{\Delta\Pi}
\newcommand{\RTau}[1]{R_{\tau, \mbox{\tiny #1}}}
\newcommand{\Nc}{N_{\mbox{\scriptsize c}}}
\newcommand{\Vud}{V_{\mbox{\scriptsize ud}}}
\newcommand{\Sew}{S_{\!\mbox{\tiny EW}}}
\newcommand{\dpew}{\delta'_{\mbox{\tiny EW}}}
\newcommand{\DeltaQCD}[1]{\Delta^{\mbox{\tiny #1}}_{\mbox{\tiny QCD}}}
\newcommand{\va}{_{\mbox{\tiny V/A}}}
\newcommand{\MTau}{M_{\tau}}
\newcommand{\ind}[2]{^{\mbox{\scriptsize $#1$}}_{\mbox{\scriptsize #2}}}
\newcommand{\cva}{\chi\va}

\newcommand{\includeplots}[2]{%
   \centerline{\includegraphics[width=72.5mm]{#1}%
   \hspace{10mm}%
   \includegraphics[width=72.5mm]{#2}}}

\begin{document}

\begin{frontmatter}

\title{Dispersive approach to QCD: $\tau$~lepton hadronic decay \\ in vector and
axial--vector channels}

\author{A.V.~Nesterenko}
\address{Bogoliubov Laboratory of Theoretical Physics,
Joint Institute for Nuclear Research \\
Joliot Curie 6, Dubna, Moscow region, 141980, Russia}
\ead{nesterav@theor.jinr.ru}

\begin{abstract}
\noindent
The dispersive approach to QCD, which extends the applicability range of
perturbation theory towards the infrared domain, is developed. This
approach properly accounts for the intrinsically nonperturbative
constraints, which originate in the low--energy kinematic restrictions on
pertinent strong interaction processes. The dispersive approach proves to
be capable of describing OPAL (update~2012) and ALEPH (update~2014)
experimental data on inclusive $\tau$~lepton hadronic decay in vector and
axial--vector channels in a self--consistent way.
\end{abstract}


\end{frontmatter}

Dispersion relations represent one of the sources of the nonperturbative
information on the hadron dynamics at low energies. In particular, the
dispersion relations render the kinematic restrictions on the pertinent
physical processes into the mathematical form and, as~a result, impose
stringent physical intrinsically nonperturbative constraints on relevant
functions. Among the latter are the hadronic vacuum polarization
function~$\Pi(q^2)$, which is defined as the scalar part of the hadronic
vacuum polarization tensor
\begin{eqnarray}
\label{PDef}
\Pi_{\mu\nu}(q^2) \!\!\!\!&=&\!\!\!\! i\!\int\!d^4x\,e^{i q x} \langle 0 |\,
T\!\left\{J_{\mu}(x)\, J_{\nu}(0)\right\} | 0 \rangle
\nonumber \\
&=& \!\!\!\!\frac{i}{12\pi^2} (q_{\mu}q_{\nu} - g_{\mu\nu}q^2) \Pi(q^2),
\end{eqnarray}
related $R(s)$~function
\begin{equation}
\label{RDef}
R(s) = \frac{1}{\pi}\, {\rm Im}\!\lim_{\varepsilon \to 0_{+}}\!
\Pi(s+i\varepsilon),
\end{equation}
which is identified with the so--called $R$--ratio of electron--positron
annihilation into hadrons, and Adler function~\cite{Adler}
\begin{equation}
\label{AdlerDef}
D(Q^2) = - \frac{d\, \Pi(-Q^2)}{d \ln Q^2},
\end{equation}
with $Q^2 = -q^2 = -s > 0$ being the spacelike kinematic variable.

The functions (\ref{PDef})--(\ref{AdlerDef}) play a crucial role in
decisive self--consistency tests of Quantum Chromodynamics~(QCD) and the
entire Standard Model, that, in turn, puts strict restrictions on possible
New Physics beyond the latter. In particular, the theoretical description
of a number of the strong interaction processes, as well as of the
hadronic contributions to precise electroweak observables is inherently
based on these functions.

The aforementioned nonperturbative constraints are properly accounted for
within dispersive approach to QCD~\cite{DQCD1, PRD88} (its preliminary
formulation was discussed in Ref.~\cite{DQCDPrelim}), which provides
unified integral representations for the functions on hand:
\begin{eqnarray}
\label{P_DQCD}
&&\hspace*{-12mm}\DP(q^2,\, q_0^2) = \DP^{(0)}(q^2,\, q_0^2)
\nonumber \\
&& \hspace*{-2.5mm}
+\!\int_{m^2}^{\infty} \rho(\sigma)
\ln\biggl(\frac{\sigma-q^2}{\sigma-q_0^2}
\frac{m^2-q_0^2}{m^2-q^2}\biggr)\frac{d\,\sigma}{\sigma}, \\[2.5mm]
\label{R_DQCD}
&&\hspace*{-12mm}R(s) = R^{(0)}(s)
+ \theta(s-m^2) \int_{s}^{\infty}\!\!
\rho(\sigma) \frac{d\,\sigma}{\sigma}, \\[2.5mm]
\label{Adler_DQCD}
&&\hspace*{-12mm}D(Q^2) = D^{(0)}(Q^2)
\nonumber \\
&& \hspace*{-2.5mm} + \frac{Q^2}{Q^2+m^2}
\int_{m^2}^{\infty} \rho(\sigma)
\frac{\sigma-m^2}{\sigma+Q^2} \frac{d\,\sigma}{\sigma}.
\end{eqnarray}
{\scriptsize
\begin{table*}[t]
\setlength{\tabcolsep}{1.4pc}
\caption{\scriptsize Values of the QCD scale parameter~$\Lambda$~[MeV]
obtained within perturbative and dispersive approaches from updated
OPAL~\cite{OPAL12} and ALEPH~\cite{ALEPH14} experimental data on inclusive
$\tau$~lepton hadronic decay (one--loop level, $n_{\mbox{\tiny f}}=3$
active flavors), see also Ref.~\cite{PRD88}.}
\label{Tab:RTau}
\vskip2mm
{\small
\centerline{\begin{tabular}{ccccc}
\hline
&
\multicolumn{2}{c}{Perturbative approach\rule[-2pt]{0pt}{10pt}}
&
\multicolumn{2}{c}{Dispersive approach}
\\
& OPAL~\cite{OPAL12}
& ALEPH~\cite{ALEPH14}
& OPAL~\cite{OPAL12}
& ALEPH~\cite{ALEPH14} \\[-0.75mm]
& ~{\scriptsize (update~2012)}~\rule[-5pt]{0pt}{5pt}
& ~{\scriptsize (update~2014)}~
& ~{\scriptsize (update~2012)}~
& ~{\scriptsize (update~2014)}~
\\ \hline
\centering
Vector channel\rule{0pt}{12.5pt}
& $445_{-230}^{+201}$
& $439_{-119}^{+110}$
& $409 \pm 53$
& $409 \pm 28$
\\[1mm]
\centering
Axial--vector channel\rule[-5.5pt]{0pt}{5.5pt}
& \multicolumn{2}{c}{no solution}
& $409 \pm 61$
& $419 \pm 33$
\\ \hline
\end{tabular}}%
\vspace*{-2.5mm}
}%
\end{table*}%
}%
In these equations $\DP(q^2\!,\, q_0^2) = \Pi(q^2) - \Pi(q_0^2)$, $m$~is
the total mass of the pertinent lightest allowed hadronic final state, and
$\theta(x)$ denotes the unit step--function [$\theta(x)=1$ if $x \ge 0$
and $\theta(x)=0$ otherwise]. The leading--order terms in
Eqs.~(\ref{P_DQCD})--(\ref{Adler_DQCD}) read~\cite{Feynman, QEDAB}:
\begin{eqnarray}
&&\hspace*{-12mm}\Delta\Pi^{(0)}(q^2,\, q_0^2) =
2\,\frac{\varphi - \tan\varphi}{\tan^3\!\varphi}
- 2\,\frac{\varphi_{0} - \tan\varphi_{0}}{\tan^3\!\varphi_{0}}, \quad \\[2.5mm]
&&\hspace*{-12mm}R^{(0)}(s) =
\theta(s - m^2)\Bigl(1-\frac{m^2}{s}\Bigr)^{\!\!3/2}, \\[2.5mm]
&&\hspace*{-12mm}D^{(0)}(Q^2) =
1 + \frac{3}{\xi}\Bigl[1 \!-\! \sqrt{1\!+\!\xi^{-1}}\,
\sinh^{-1}\!\bigl(\xi^{1/2}\bigr)\Bigr],
\end{eqnarray}
whereas $\rho(\sigma)$ denotes the spectral density
\begin{eqnarray}
\label{RhoGen}
&&\hspace*{-12mm}
\rho(\sigma) = \frac{1}{\pi} \frac{d}{d\,\ln\sigma}\,
\mbox{Im}\lim_{\varepsilon \to 0_{+}} p(\sigma-i\varepsilon)
\nonumber \\[1.5mm]
&&\hspace*{-7.5mm}
= - \frac{d\, r(\sigma)}{d\,\ln\sigma}
= \frac{1}{\pi}\, \mbox{Im}\lim_{\varepsilon \to 0_{+}}
d(-\sigma-i\varepsilon).
\end{eqnarray}
Here $\sin^2\!\varphi = q^2/m^2$, $\sin^2\!\varphi_{0} = q^{2}_{0}/m^2$,
$\xi=Q^2/m^2$, and $p(q^2)$, $r(s)$, $d(Q^2)$ stand for the strong
corrections to functions~(\ref{PDef}), (\ref{RDef}), (\ref{AdlerDef}),
respectively (see Refs.~\cite{DQCD1, PRD88} for the details).

It is worth mentioning that the derivation of representations
(\ref{P_DQCD})--(\ref{Adler_DQCD}) involves no phenomenological
assumptions. The Adler function~(\ref{Adler_DQCD}) agrees with
corresponding experimental prediction in the entire energy
range~\cite{DQCD1, DQCD2} (the studies of~$D(Q^2)$ can also be found in
Refs.~\cite{Maxwell, PeRa, Kataev, MSS, Cvetic, BJ, Fischer1, Fischer2}),
the functions (\ref{P_DQCD})--(\ref{Adler_DQCD}) comply with the results
obtained in Refs.~\cite{PRL99PRD77, RCTaylor}, and the hadronic vacuum
polarization function~(\ref{P_DQCD}) conforms with relevant lattice
simulation data~\cite{Lattice, Prep}.

The unambiguous method to restore the complete expression for the spectral
density~$\rho(\sigma)$~(\ref{RhoGen}) is still far from being feasible
(for a discussion of this issue see, e.g., Refs.~\cite{PRD62, DQCD4}).
Nonetheless, the perturbative contribution to~$\rho(\sigma)$ can be
calculated by making use of the perturbative expression for either of the
strong corrections to the functions on hand (see, e.g., Ref.~\cite{CPC})
\begin{eqnarray}
\label{RhoPert}
&&\hspace*{-12mm}
\rho\ind{}{pert}(\sigma) = \frac{1}{\pi} \frac{d}{d\,\ln\sigma}\,
\mbox{Im}\lim_{\varepsilon \to 0_{+}} p\ind{}{pert}(\sigma-i\varepsilon)
\nonumber \\[1.5mm]
&&\hspace*{-10mm}
= \! - \frac{d\, r\ind{}{pert}(\sigma)}{d\,\ln\sigma}
\!=\! \frac{1}{\pi}\, \mbox{Im}\lim_{\varepsilon \to 0_{+}}
d\ind{}{pert}(-\sigma-i\varepsilon).
\end{eqnarray}
In this paper the model~\cite{PRD88} for the spectral density will be
employed:
\begin{equation}
\label{RhoMod}
\rho(\sigma) = \frac{4}{\beta_{0}}\frac{1}{\ln^{2}(\sigma/\Lambda^2)+\pi^2} +
\frac{\Lambda^2}{\sigma}.
\end{equation}
The first term on the right--hand side of Eq.~(\ref{RhoMod}) is the
one--loop perturbative contribution, whereas the second term represents
intrinsically nonperturbative part of the spectral density, see
paper~\cite{PRD88}.

It is worthwhile to note that in the massless limit ($m=0$) for the case
of perturbative spectral function [$\rho(\sigma) = \mbox{Im}\;
d\ind{}{pert}(-\sigma - i\,0_{+})/\pi$] two representations~(\ref{R_DQCD})
and~(\ref{Adler_DQCD}) become identical to those of the so--called
analytic perturbation theory~\cite{APT} (see also Refs.~\cite{APT1, APT2,
APT3, APT4, APT5, APT6, APT7, APT8}). However, it is essential to keep the
value of the hadronic production threshold~$m$ nonvanishing, since the
massless limit loses some of the nonperturbative constraints, which
relevant dispersion relations impose on the functions on hand, see
paper~\cite{PRD88} and references therein for the details.

\begin{figure*}[t]
\includeplots{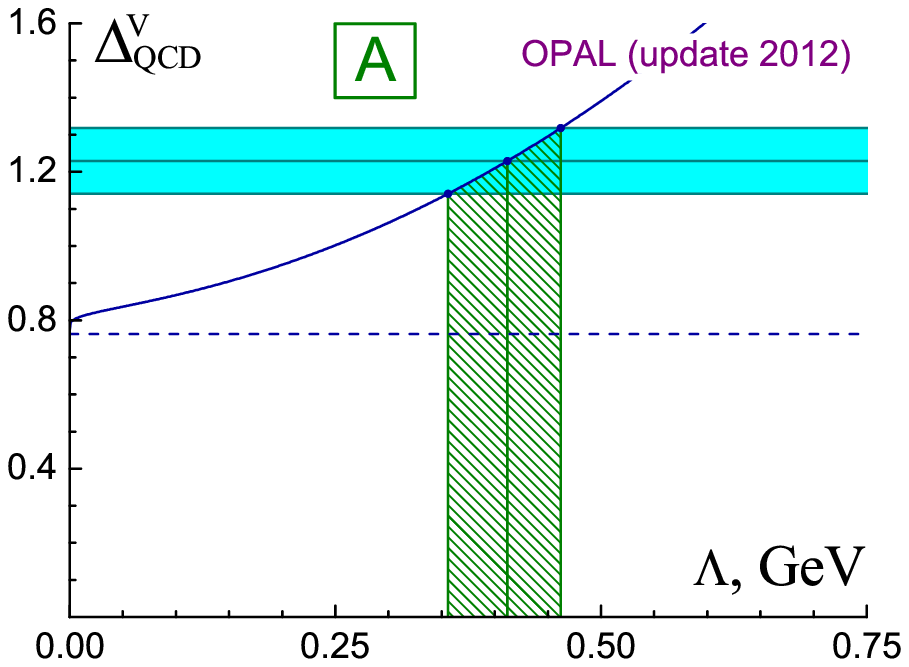}{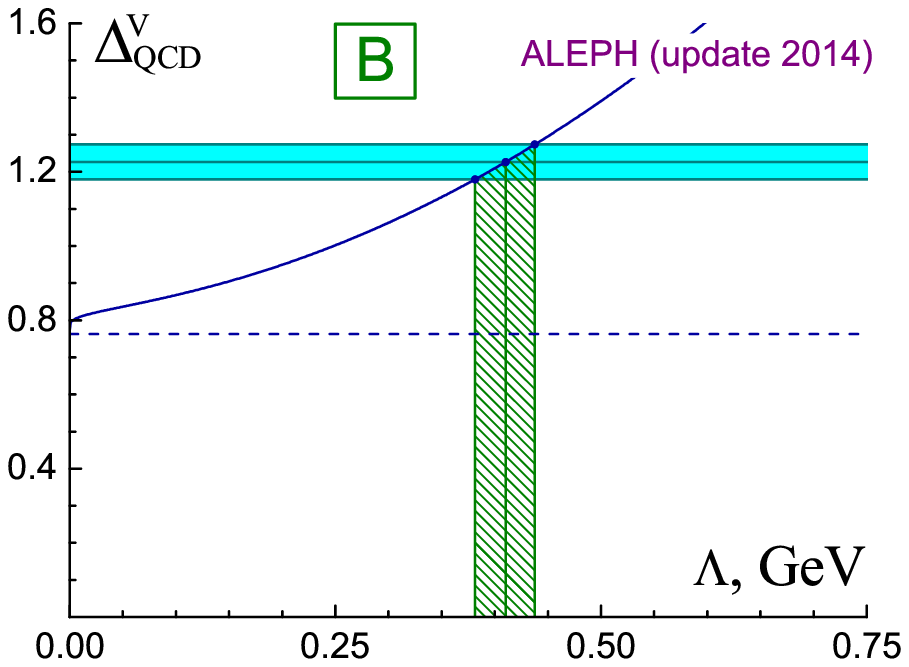}
\vskip5mm
\includeplots{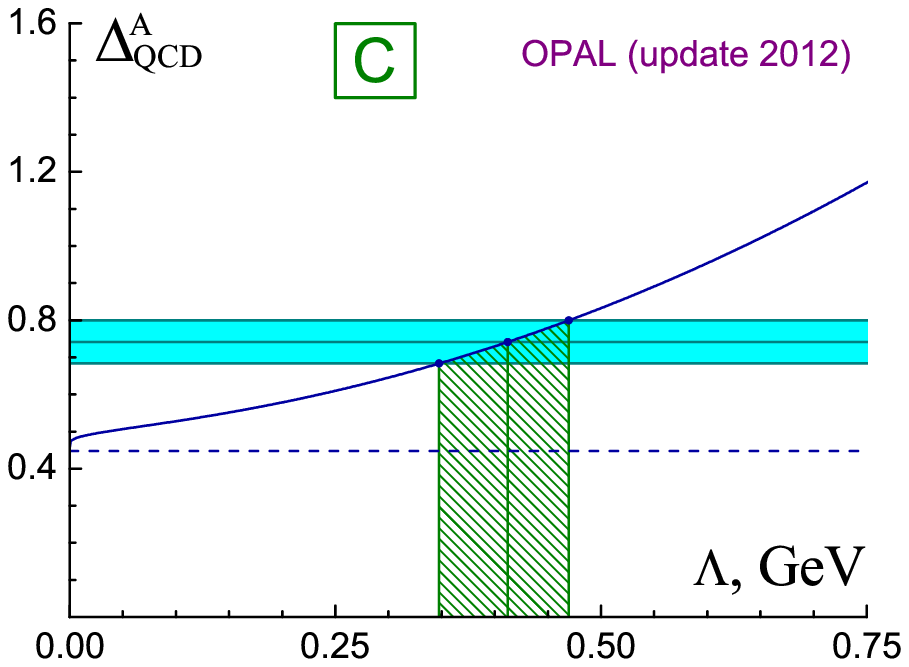}{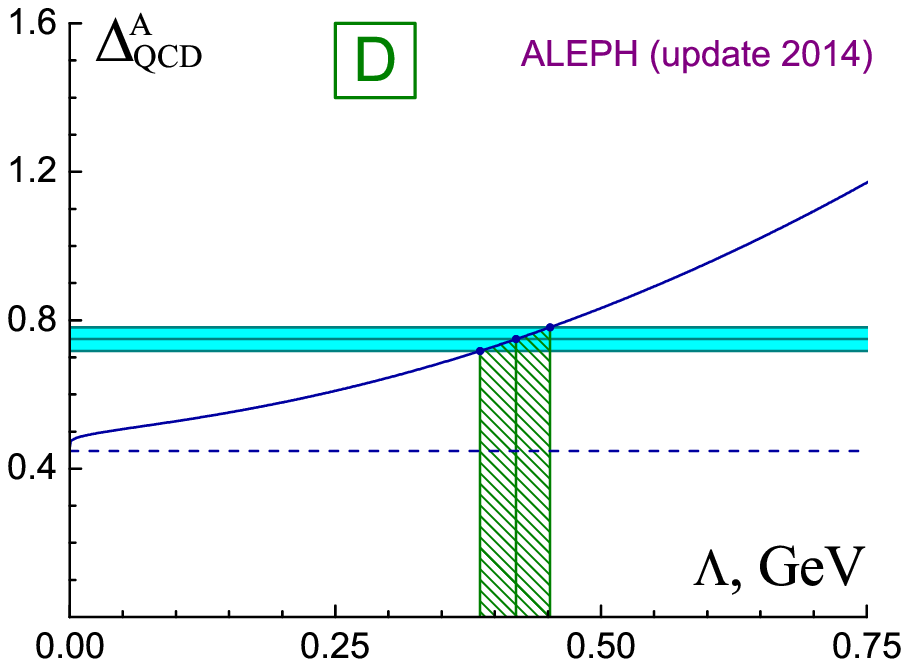}
\caption{\scriptsize Comparison of the
expression~$\DeltaQCD{V/A}$~(\ref{DeltaQCD}) (solid curves) with relevant
experimental data (horizontal shaded bands). Vertical dashed bands denote
solutions for the QCD scale parameter~$\Lambda$. The plots A,~C and B,~D
correspond to experimental data~\cite{OPAL12} and~\cite{ALEPH14},
respectively.}
\label{Plot:RTauDQCD}
\end{figure*}

\smallskip

The dispersive approach has been successfully applied to the study of the
inclusive $\tau$~lepton hadronic decay in vector~(V) and axial--vector~(A)
channels. Specifically, the theoretical expression for the pertinent
experimentally measurable quantity reads
\begin{equation}
\label{RTauGen}
\RTau{V/A}^{\tiny{J=1}} = \frac{\Nc}{2}\, |\Vud|^2\,\Sew\,
\Bigl(\DeltaQCD{V/A} + \dpew \Bigr),
\end{equation}
where $\Nc=3$ is the number of colors, $|\Vud| = 0.97425 \pm 0.00022$ is
Cabibbo--Kobayashi--Maskawa matrix element~\cite{PDG2012}, $\Sew = 1.0194
\pm 0.0050$ and $\dpew = 0.0010$ stand for the electroweak
corrections~\cite{EWF}, and
\begin{eqnarray}
\label{DeltaQCDDef}
&&\hspace*{-10mm}
\DeltaQCD{V/A} = \frac{2}{\pi}\int_{m\va^2}^{\MTau^2}\!
\biggl(1-\frac{s}{\MTau^2}\biggr)^{\!\!2}\biggl(1+2\frac{s}{\MTau^2}\biggr)
\nonumber \\
&&\times\,\mbox{Im}\,\Pi^{\mbox{\tiny V/A}}(s+i0_{+})\, \frac{d s}{\MTau^2}
\end{eqnarray}
denotes the hadronic contribution, see Ref.~\cite{BNPPDP}.
In~Eq.~(\ref{DeltaQCDDef}) $\MTau \simeq 1.777\,$GeV~\cite{PDG2012} is the
mass of $\tau$~lepton and $m\va$~stands for the total mass of the lightest
allowed hadronic decay mode of $\tau$~lepton in the corresponding channel.

It is worth noting that the description of the inclusive $\tau$~lepton
hadronic decay within perturbative approach completely leaves out the
effects due to the nonvanishing hadronic production threshold.
Additionally, this approach suffers from its inherent difficulties, such
as the infrared unphysical singularities. These facts eventually lead to
the identity of the perturbative predictions for
functions~(\ref{DeltaQCDDef}) in vector and axial--vector channels
($\Delta\ind{\mbox{\tiny V}}{pert}=\Delta\ind{\mbox{\tiny A}}{pert}$, that
contradicts experimental data~\cite{OPAL12, ALEPH14}) and the failure of
the perturbative approach to describe the experimental data on the
inclusive semileptonic branching ratio in axial--vector channel, see
Table~\ref{Tab:RTau}. Note also that for vector channel perturbative
approach yields two solutions for the QCD scale parameter~$\Lambda$, one
of which is commonly discarded, see paper~\cite{PRD88} and references
therein.

The inclusive $\tau$~lepton hadronic decay was also studied within
massless analytic perturbation theory and a number of its
modifications~\cite{MSS, TauAPT1, TauAPT2}. However, these papers
basically deal either with the total sum of vector and axial--vector
terms~(\ref{RTauGen}) or with the vector term only.

In the framework of the dispersive approach the hadronic
contribution~(\ref{DeltaQCDDef}) to the inclusive semileptonic
branching ratio can be represented in the following form:
\begin{eqnarray}
\label{DeltaQCD}
&&\hspace*{-12mm}
\DeltaQCD{V/A} = 3\,g_{1}\biggl(\frac{\cva}{2}\biggr)\sqrt{1-\cva}
\nonumber \\[1mm]
&&\hspace*{-7.5mm}
- 3\,g_{2}\biggl(\frac{\cva}{4}\biggr)
\ln\biggl(\sqrt{\cva^{-1}}+\sqrt{\cva^{-1}-1}\biggr)
\nonumber \\[1mm]
&&\hspace*{-7.5mm}
+ \int_{m\va^{2}}^{\infty}\!G\Bigl(\frac{\sigma}{M_{\tau}^{2}}\Bigr)\,
\rho(\sigma)\,\frac{d \sigma}{\sigma}\,,
\end{eqnarray}
where
$G(x) = g(x)\,\theta(1-x) + g(1)\,\theta(x-1) - g(\chi\va)$,
$g(x) = x (2 - 2x^2 + x^3)$,
$\chi\va = m\va^{2}/\MTau^{2}$, $m_{\mbox{\tiny V}}^{2} \simeq
0.075\,\mbox{GeV}^2$, $m_{\mbox{\tiny A}}^{2} \simeq 0.288\,\mbox{GeV}^2$,
spectral density~$\rho(\sigma)$ is specified in Eq.~(\ref{RhoMod}), and
\begin{eqnarray}
g_{1}(x) \!\!\!\!&=&\!\!\!\! \frac{1}{3} + 4x -\frac{5}{6}x^2 + \frac{1}{2}x^3, \\[1.5mm]
g_{2}(x) \!\!\!\!&=&\!\!\!\! 8x(1 + 2x^2 - 2x^3),
\end{eqnarray}
see papers~\cite{PRD88, DQCD4, DQCD3} and references therein. The
juxtaposition of the obtained result~(\ref{DeltaQCD}) with recently
updated OPAL~\cite{OPAL12} and ALEPH~\cite{ALEPH14} experimental data is
presented in Fig.~\ref{Plot:RTauDQCD} and the corresponding values of the
QCD scale parameter~$\Lambda$ are listed in Table~\ref{Tab:RTau}. As one
can infer from Fig.~\ref{Plot:RTauDQCD}, the dispersive approach is
capable of describing the experimental data~\cite{OPAL12, ALEPH14} on
inclusive $\tau$~lepton hadronic decay in vector and axial--vector
channels. The obtained values of the QCD scale parameter~$\Lambda$ appear
to be nearly identical in both channels, that testifies to the
self--consistency of the developed approach.

\medskip

The author is grateful to D.~Boito, S.~Brodsky, A.~Kataev, B.~Malaescu,
and H.~Wittig for the stimulating discussions and useful comments.

\end{document}